\documentclass[conference]{IEEEtran}
\IEEEoverridecommandlockouts
\usepackage{cite}
\usepackage{amsmath,amssymb,amsfonts}
\usepackage{algorithmic}
\usepackage{graphicx}
\usepackage{textcomp}
\usepackage[table]{xcolor}
\usepackage{hhline}

\usepackage{url}
\usepackage{multirow}

\usepackage[caption=false,labelformat=simple]{subfig}

\usepackage{hyperref}
    
\begin{document}

\title{Should We Worry About Interference in \\ Emerging Dense NGSO Satellite Constellations?}

\author{\IEEEauthorblockN{Christophe Braun, Andra M. Voicu, Ljiljana Simi\'c and Petri M\"ah\"onen}
Institute for Networked Systems \\
RWTH Aachen University\\
Email: \{cbr, avo, lsi, pma\}@inets.rwth-aachen.de
}

\maketitle

\begin{abstract}
Many satellite operators are currently planning to deploy non-geostationary-satellite orbit (NGSO) systems for broadband communication services in the Ku-, Ka-, and V-band, where some of them have already started launching.
Consequently, new challenges are expected for inter-system satellite coexistence due to the increase in the interference level and the complexity of the interactions resulting from the heterogeneity of the constellations. 
This is especially relevant for the Ku-band, where the NGSO systems are most diverse and existing geostationary-satellite orbit (GSO) systems, which often support critical services, must be protected from interference.
It is thus imperative to evaluate the impact of mutual inter-system interference, the efficiency of the basic interference mitigation techniques, and whether regulatory intervention is needed for these new systems. 
We conduct an extensive study of inter-satellite coexistence in the Ku-band, where we consider all recently proposed NGSO and some selected GSO systems. 
Our throughput degradation results suggest that existing spectrum regulation may be insufficient to ensure GSO protection from NGSO interference, especially due to the high transmit power of the low Earth orbit (LEO) Kepler satellites. 
This also results in strong interference towards other NGSO systems, where traditional interference mitigation techniques like \emph{look-aside} may perform poorly.    
Specifically, \emph{look-aside} can be beneficial for large constellations, but detrimental for small constellations. 
Furthermore, we confirm that band-splitting among satellite operators significantly degrades throughput, also for the Ku-band.  
Our results overall show that the complexity of the inter-satellite interactions for new NGSO systems is too high to be managed via simple interference mitigation techniques. 
This means that more sophisticated engineering solutions, and potentially even more strict regulatory requirements, will be needed to ensure coexistence in emerging, dense NGSO deployments.    
\end{abstract}


\section{Introduction}

With the ongoing demand for broadband services, network operators have been diversifying the range of deployed wireless technologies and their applications.
In this context, satellite communication systems are being increasingly used for e.g. backhaul infrastructure for on-board wireless connectivity in airplanes~\cite{E.Dincemphetal.2017} and integration with cellular 5G to offer terrestrial broadband services~\cite{Giambene2018}.
Recently, a large number of satellite operators have applied to the US spectrum regulator FCC for permission to launch new non-geostationary-satellite orbit (NGSO) systems~\cite{scheds}.
These systems target spectrum bands in the range of 10-52~GHz, i.e. the Ku-, Ka-, and V-band~\cite{ITU-R2015}, and some have already been approved and started launching, e.g. SpaceX~\cite{SpaceX2019}.

NGSO satellite deployments are thus expected to undergo significant densification compared to existing systems. This will lead to challenging inter-satellite coexistence cases in shared bands due to (i)~the increase in the interference level, and (ii)~the expected high heterogeneity of the NGSO satellite systems, for which the current spectrum regulation is very permissive, as e.g. enforced by the FCC in the US. 
In this dynamic emerging satellite deployment landscape, it is important to understand the interference interactions among different NGSO systems and to what extent regulatory intervention may be needed to ensure NGSO-NGSO coexistence.  

The most challenging inter-satellite coexistence cases are expected in the Ku-band, due to the very heterogeneous systems in terms of numbers of satellites (i.e. tens to thousands) and geometric orbit properties (e.g. circular, elliptical) that are set to operate in this band. 
Moreover, existing geostationary-satellite orbit (GSO) systems also operate in the Ku-band and must be protected from interference by NGSO systems, as enforced by regulation in the US~\cite{CodeofFederalRegulations}, where applications for a license for these new systems have been initiated.
Given this high heterogeneity and uncertainty about the exact parameters of emerging deployments, it is not yet clear whether traditional satellite interference mitigation techniques and existing regulatory requirements are sufficient to ensure coexistence with the new NGSO systems.

Although there is some prior work on the impact of interference and mitigation techniques for satellite systems, most authors considered only interference between GSO-NGSO systems, e.g.~\cite{InLineMitTech, coexsepangle}. Moreover, these works considered NGSO legacy deployments with few satellites, where NGSO-NGSO interference was not an issue, as expected for emerging constellations. 
Importantly, NGSO-NGSO inter-system interference has been largely unaddressed in literature, with the notable exception of~\cite{impactsatdiv, TPRC}. The authors in~\cite{impactsatdiv} considered only two low Earth orbit (LEO) constellations and one medium Earth orbit (MEO) constellation, with a satellite diversity technique to mitigate interference. This is different to the emerging NGSO satellite deployments, where many more systems with different design parameters are expected to coexist. 
Consequently, it is not clear whether the interference mitigation technique analysed in~\cite{impactsatdiv} is efficient for dense deployments.  
The authors in~\cite{TPRC} conducted an extensive study on the impact of  NGSO-NGSO co-channel interference in terms of throughput for new NGSO constellations with interference mitigation techniques like look-aside and band-splitting in the Ka- and V-band. 
However, in these bands there are no GSO satellite systems and no NGSO systems with elliptical or geosynchronous orbits.
By contrast, in the Ku-band, we expect such more challenging coexistence cases. 
Consequently, it is imperative to thoroughly analyse inter-system coexistence for satellite deployments in the Ku-band, due to the highly heterogeneous NGSO constellation properties and NGSO-GSO interference interactions.  
 
In this paper we consider inter-satellite coexistence in the Ku-band, conducting an extensive study on the impact of both NGSO-NGSO and NGSO-GSO co-channel interference on throughput.
We adopt the methodology in~\cite{TPRC}, which we extend to incorporate GSO systems and more diverse NGSO architectures, i.e. highly elliptical orbit (HEO) and geosynchronous constellations.  
We consider various traditional interference mitigation techniques like \emph{look-aside} and \emph{band-splitting} for several ground station locations in the US and Europe. 
Our work is thus the first comparative analysis on coexisting NGSO-NGSO and NGSO-GSO systems for a diverse set of scenarios, enabling us to derive insights about the efficiency of interference mitigation techniques with respect to different orbits, transceiver parameters, and ground station locations on Earth.  

\begin{figure}[!t]
\begin{center}
\centering
  \includegraphics [width = 0.55\columnwidth] {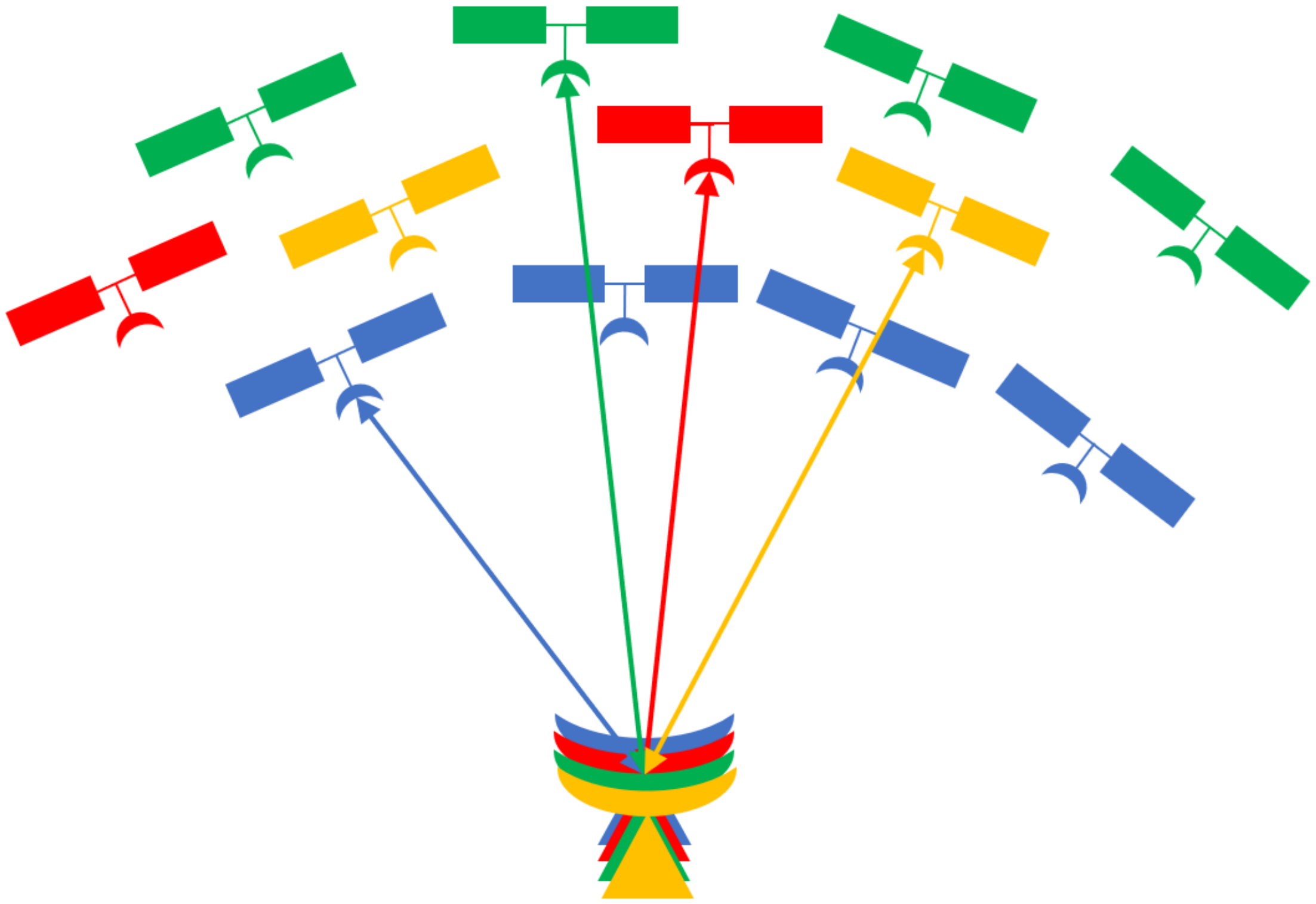}
  \caption{Illustration of inter-satellite co-channel interference among different systems in the downlink, showing the GSO system (green) and three NGSO systems (yellow, red, and blue). All ground stations are co-located on Earth and each of them communicates with a single available satellite from its corresponding constellation. For each given link, all other links (in different colors) are interfering links.}\label{fig:scenariosngso}
\end{center}
\end{figure}

Our results indicate that current spectrum regulation may be insufficient to ensure efficient GSO protection from NGSO interference. Furthermore, the most promising traditional NGSO-NGSO interference mitigation technique is look-aside, however, its performance is highly sensitive to constellation properties and the locations of ground stations. This suggests that more sophisticated engineering solutions and potentially more strict regulatory constraints are required to ensure coexistence of emerging NGSO deployments.

The remainder of this paper is structured as follows. Section~\ref{sysmod} presents the system model. Section~\ref{PerEvaMet} details the simulation setup. Section~\ref{results} presents and discusses the throughput results. Section~\ref{conclusions} concludes the paper.

\section{System Model}
\label{sysmod}

This section presents the system model used to study the impact of interference from NGSO satellite systems. We first elaborate the considered interference types and scenarios in Section~\ref{interfTypes}. We then present the considered interference mitigation techniques in Section~\ref{IntMitTech}, our evaluation metric in Section~\ref{EvalMet}, and the satellite constellations in Section~\ref{satconst}.

\subsection{Interference Types \& Scenarios}
\label{interfTypes}

\begin{figure}[!t]
	\centering	
	\subfloat[Look-aside]{\includegraphics [width = 0.4\columnwidth] {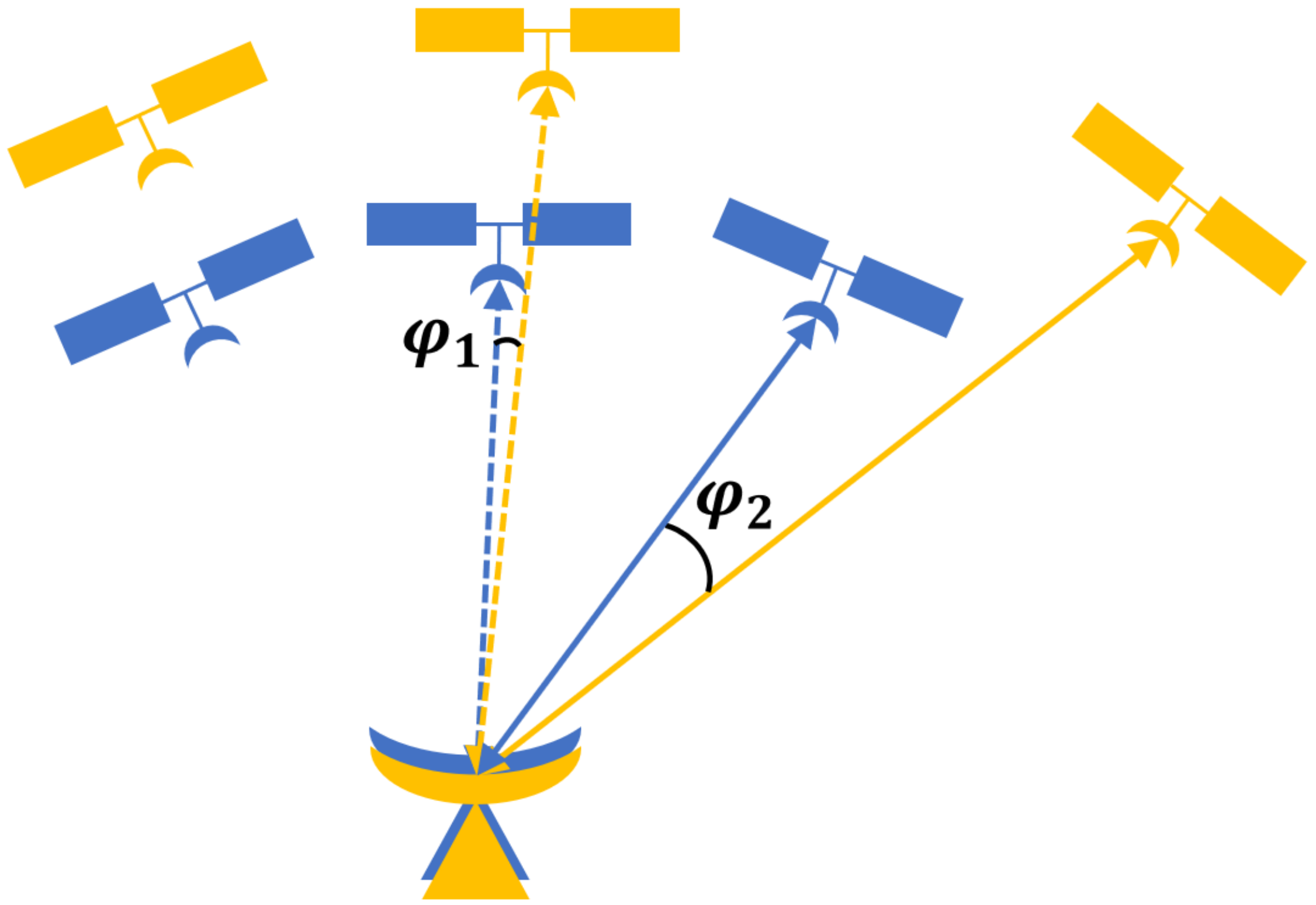}\label{fig:sclookasdide}}
	~
	\subfloat[GSO protection] {\includegraphics [width = 0.4\columnwidth] {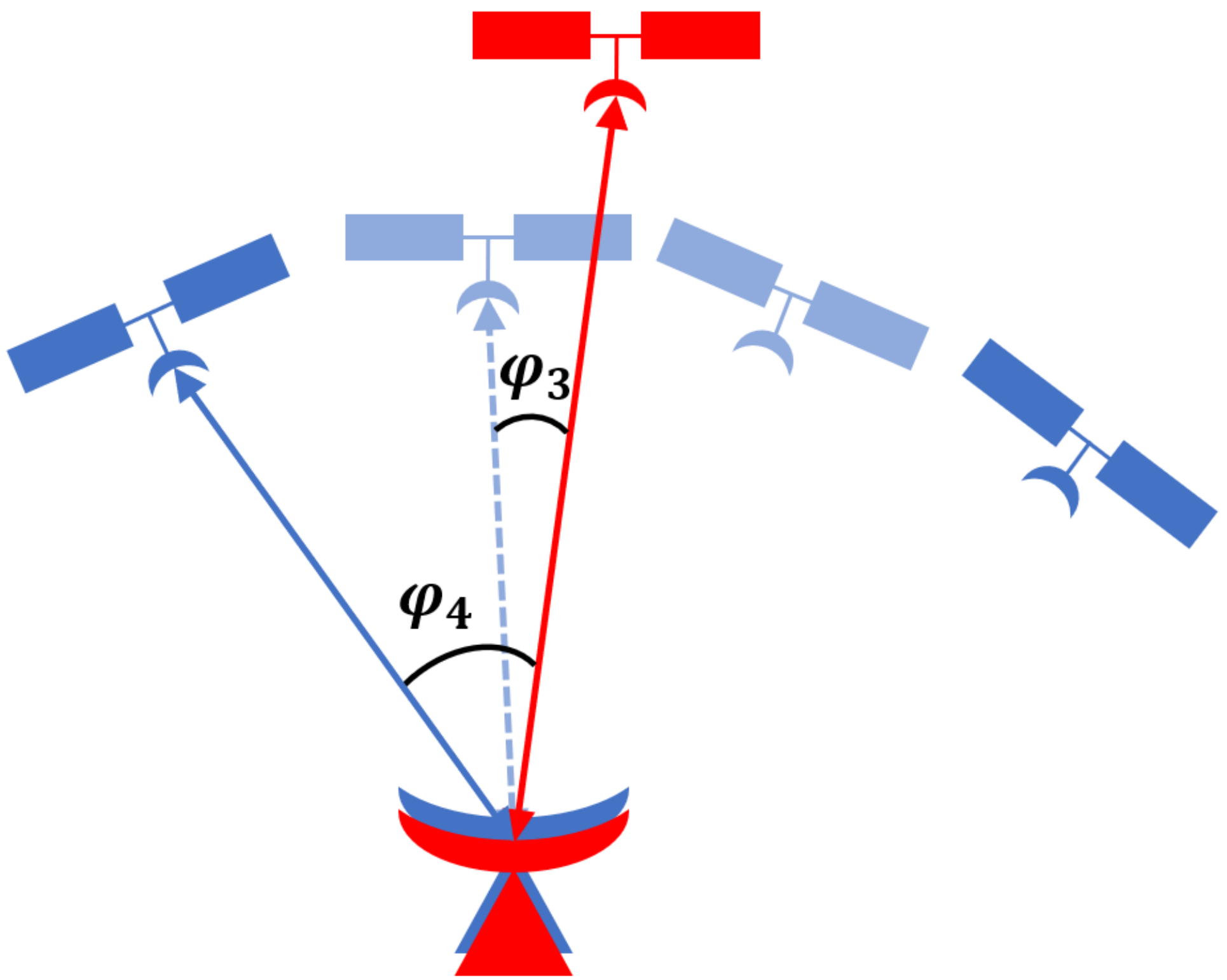} \label{fig:scgsoprot}}	
	\caption{Illustration of the satellite diversity interference-mitigation techniques (a)~look-aside between NGSO systems (yellow and blue) and (b)~GSO protection between an NGSO system (blue) and a GSO system (red). Angles $\varphi_1$ and $\varphi_3$ occur when no interference mitigation is applied, whereas $\varphi_2>5^{\circ}$ and $\varphi_4>30^{\circ}$ occur with the two respective interference mitigation techniques.}
		\label{fig:interfmitig}
\end{figure}

\begin{figure*}[!t]
	\centering	
	\subfloat[OneWeb LEO]{\includegraphics [width = 0.15\linewidth] {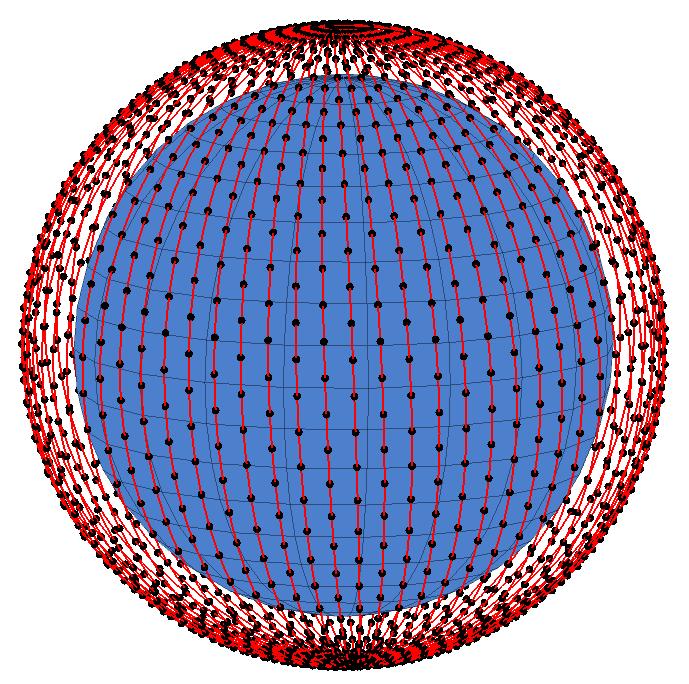}\label{fig:onewebleoorb}}
	~	
	\subfloat[SpaceX LEO] {\includegraphics [width = 0.15\linewidth] {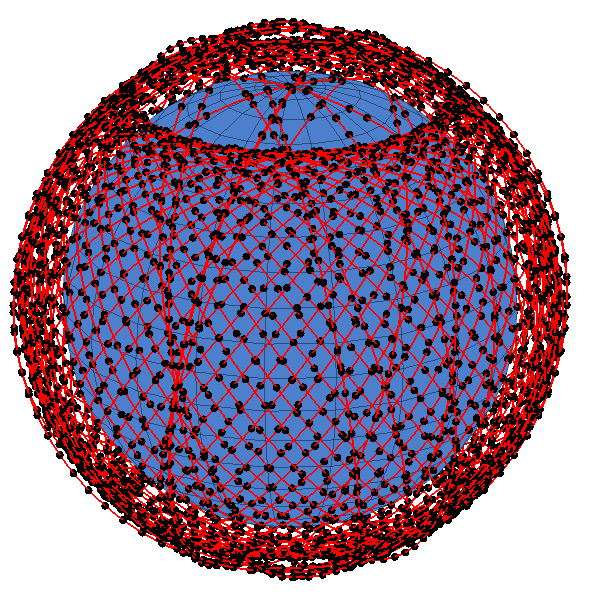} \label{fig:spacexorb}}
	~
	\subfloat[Kepler LEO]{\includegraphics [width = 0.15\linewidth] {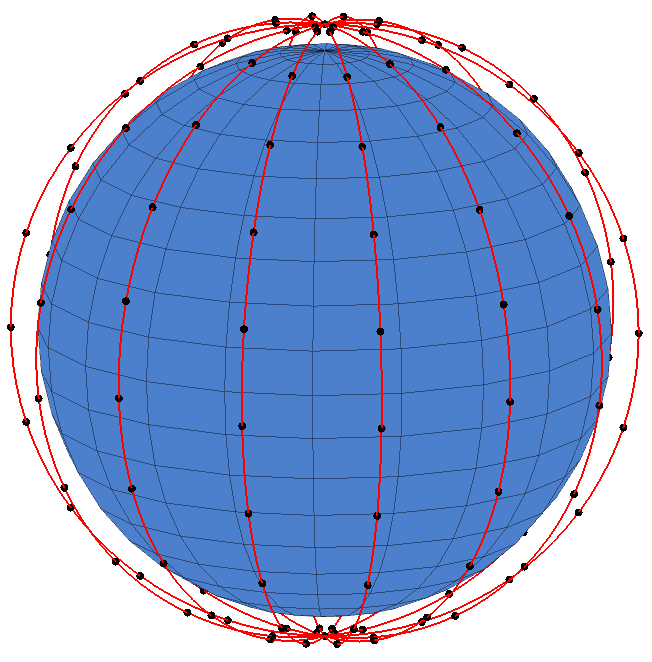}\label{fig:keplerorb}}
	~
	\subfloat[Theia LEO]{\includegraphics [width = 0.15\linewidth] {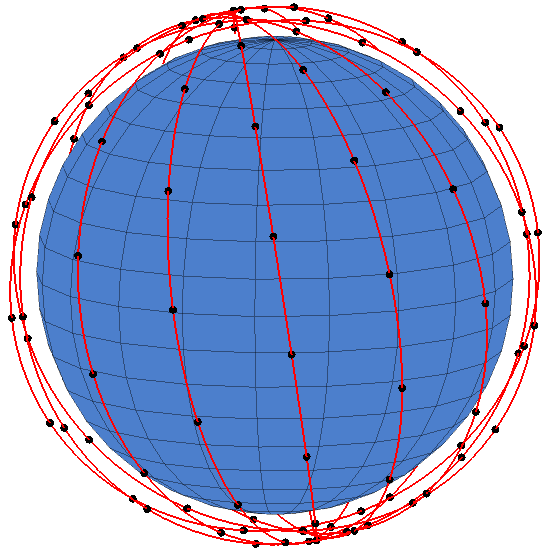}\label{fig:theiaorb}}
	~
	\subfloat[OneWeb MEO]{\includegraphics [width = 0.15\linewidth] {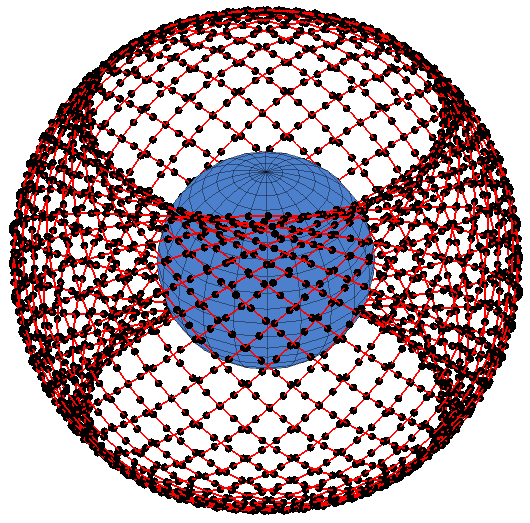}\label{fig:onewebmeoorb}}
	\\
	\subfloat[Karousel Geosynchronous]{\includegraphics [width = 0.25\linewidth] {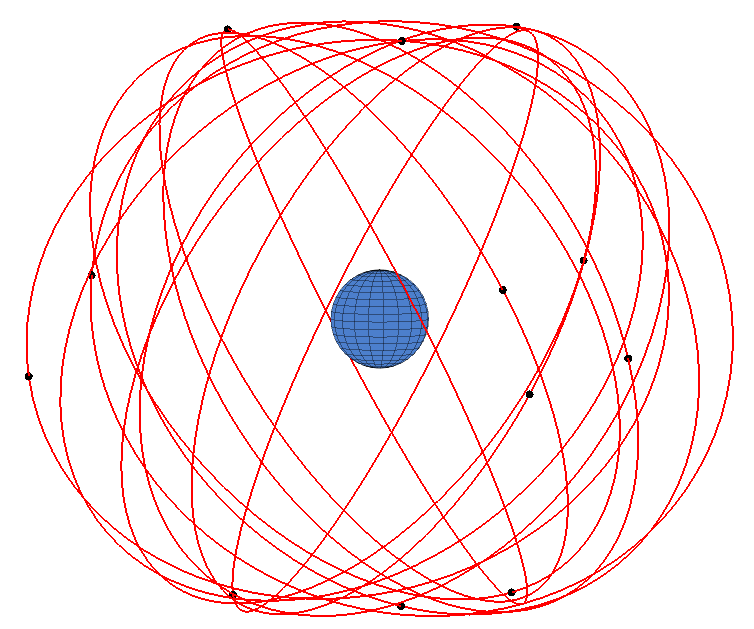}\label{fig:karouselorb}}
	~
	\subfloat[Space Norway HEO] {\includegraphics [width = 0.17\linewidth] {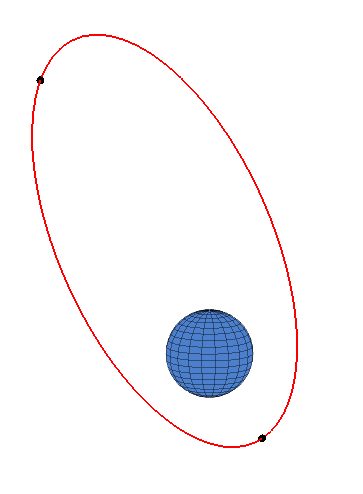}\label{fig:spacenororb}}
	~
	\subfloat[NSS HEO] {\includegraphics [width = 0.15\linewidth] {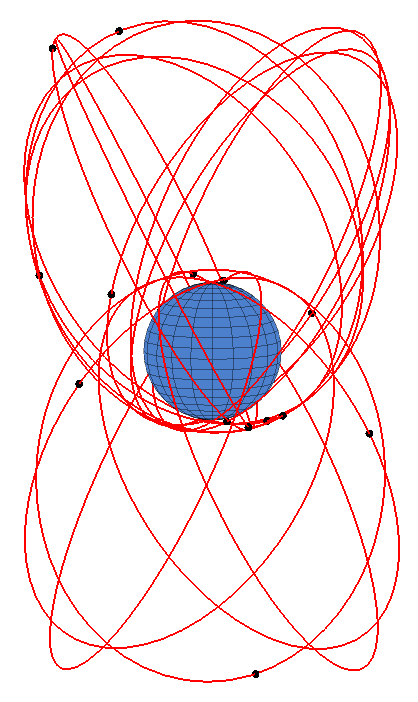}\label{fig:nssorb}}	
	~
	\subfloat[GSO] {\includegraphics [width = 0.25\linewidth] {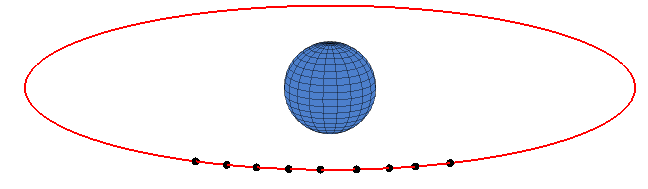}\label{fig:gsoorb}}
	\caption{Illustration of all considered satellite constellations, showing the Earth (blue), the satellite orbits (red), and the satellites (black).}
		\label{fig:lmeoconst}
\end{figure*}

We focus on co-channel co-polarized\footnote{We note that some satellite systems use cross-polarization to distinguish different co-channel transmissions and thus to increase the link capacity~\cite{SatComSysEng}. This is outside the scope of this paper.} interference among NGSO-NGSO and among NGSO-GSO satellite systems operating in the Ku-band in the downlink. This is illustrated in Fig.~\ref{fig:scenariosngso} for different systems in terms of altitude, number of available satellites covering a given Earth location, and elevation angle of the selected satellite.
There is one ground station per NGSO constellation and one ground station per GSO satellite, where all stations are co-located on Earth. We thus consider the worst-case interference where the directional antennas of an interfering system are pointed towards the victim system, if no interference mitigation technique is applied.
However, we expect our results to be relevant also for practical separation distances, since~\cite{TPRC} reported that ground stations of NGSO systems operating in the Ka- and V-band could be considered as co-located for spatial separations of up to 20~km, and as nearly co-located for a separation of 100~km. 

Each NGSO ground station communicates with a single (available) satellite from its corresponding constellation and we consider only inter-system interference; managing intra-system interference is a less challenging case, due to the existence of a single operator that owns and configures the entire system. 
For GSO satellites, we also model GSO-GSO interference, since every considered satellite is deployed by a different operator; however, mitigating GSO-GSO interference is outside the scope of this paper.
In~\cite{Braun2019} results were presented also for the uplink, but since the parameters of the ground stations are largely not yet specified by the NGSO operators, most of whom have not yet applied for licenses for ground stations, we omit these results here. 
  
We consider three different scenarios: (i)~\emph{\textbf{baseline}}, where each satellite system operates individually without inter-system interference; (ii)~\emph{\textbf{NGSO-NGSO interference}}, where there are only NGSO systems; and (iii)~\emph{\textbf{NGSO-GSO interference}}, where multiple NGSO systems and one GSO system coexist. 
For the last two scenarios, inter-system interference is considered both for pairs of systems, and as aggregate from all other systems.
In addition, we consider two sets of transceiver parameters for the NGSO systems: the \emph{\textbf{original}} parameters as proposed by the operators in~\cite{scheds}, and \emph{\textbf{tuned}} parameters that we adjust to achieve better harmonization between systems at the same altitude (\emph{cf.} Appendix).

\subsection{Interference Mitigation Techniques}
\label{IntMitTech}

\subsubsection{Look-Aside}
This is a distributed satellite diversity technique which imposes a minimum separation angle between the link of a victim satellite system and all interfering links from other systems. We assume this angle to be at least 5°, which was found in~\cite{TPRC} to achieve a good tradeoff between the performance of large and small constellations.
Fig.~\ref{fig:sclookasdide} shows the link selection between a satellite and a ground station without interference mitigation, and for look-aside. Without interference mitigation, the yellow ground station selects a yellow satellite, such that the separation angle between the victim and the closest interfering satellite in blue is $\varphi_1$. When the look-aside mitigation technique is applied, the yellow ground station selects a yellow satellite separated by $\varphi_2>5^{\circ}$ from the closest interfering satellite in blue. The link selection process is discussed in more detail in Section~\ref{linksel}.
\subsubsection{Band-Splitting}
This was introduced by the FCC in~\cite{fccrule}. Whenever the noise temperature at the victim or interfering receiver is increased by 6\%, band-splitting must be triggered, so that the available bandwidth is equally split between the operators. We note that this technique aims at managing strong inter-system interference if no other engineering solutions are found and is expected to have a strong impact on the throughput, such that other interference-mitigation techniques may be preferred in practice. 
\subsubsection{GSO Protection Technique}
In order to protect GSO from NGSO systems as required by the FCC~\cite{CodeofFederalRegulations} and \mbox{ITU-R}~\cite{radioreg}, we apply a satellite diversity technique similar to look-aside, as predominantly proposed in~\cite{scheds}. Specifically, an NGSO satellite is allowed to communicate with its ground station only if there is a separation angle of at least 30° from all GSO satellites. Fig. \ref{fig:scgsoprot} illustrates an example of link selection when the GSO protection technique is applied. The light blue NGSO satellites are too close to the GSO satellite ($\varphi_3<30^{\circ}$) and they are thus not allowed to communicate with the blue NGSO ground station. Only the dark blue satellites are allowed to communicate with the blue ground station, where $\varphi_4>30^{\circ}$.

\subsection{Evaluation Metric}
\label{EvalMet}

We consider the throughput degradation $\Delta R$ with respect to a fixed reference value as our evaluation metric, consistent with~\cite{fccrisk}. First, the carrier-to-interference-plus-noise ratio $C/(I+N)$ is determined at the victim receiver~\cite{TPRC} as
\begin{equation}
C/(I+N) = -10\log_{10}\Bigg((N/C)_{lin}+\sum_{i=1}^N(I_i/C)_{lin}\Bigg)[dB],
\label{eq:cinfinal}
\end{equation} 
where $(N/C)_{lin}$ and $(I_i/C)_{lin}$ are the multiplicative inverses of the carrier-to-noise ratio $C/N$ from the victim satellite system and the carrier-to-interference ratio from the $i$-th interfering satellite system $C/I_i$, in linear scale. Terms $C/N$ and $C/I_i$ depend on the respective free-space path loss $L_{FS}$ and atmospheric attenuation $A_T$~\cite{TPRC}.

We assume that adaptive coding and modulation is implemented, where the actual spectral efficiency $SE_{Act}$ that corresponds to $C/(I+N)$ is determined according to the specifications of the DVB-S2X standard~\cite{dvbs}. Finally, $\Delta R$ is calculated as
\begin{equation}
\Delta R = 1-\frac{BW_{Act}\times SE_{Act}}{BW_{Full}\times SE_{Ref}},
\label{eq:dtp}
\end{equation}  
where $BW_{Act}$ is the actual bandwidth, $BW_{Full}$ is the full available bandwidth, and $SE_{Ref}$ is the reference spectral efficiency equal to the maximum spectral efficiency of the DVB-S2X standard of 5.90~bits/s/Hz. We note that $BW_{Act}$=$BW_{Full}$, except for triggered band-splitting~\cite{TPRC}.

\subsection{Satellite Constellations \& Earth Locations}\label{satconst}

\begin{figure}[!t]
\begin{center}
\centering
  \includegraphics [scale = 0.23] {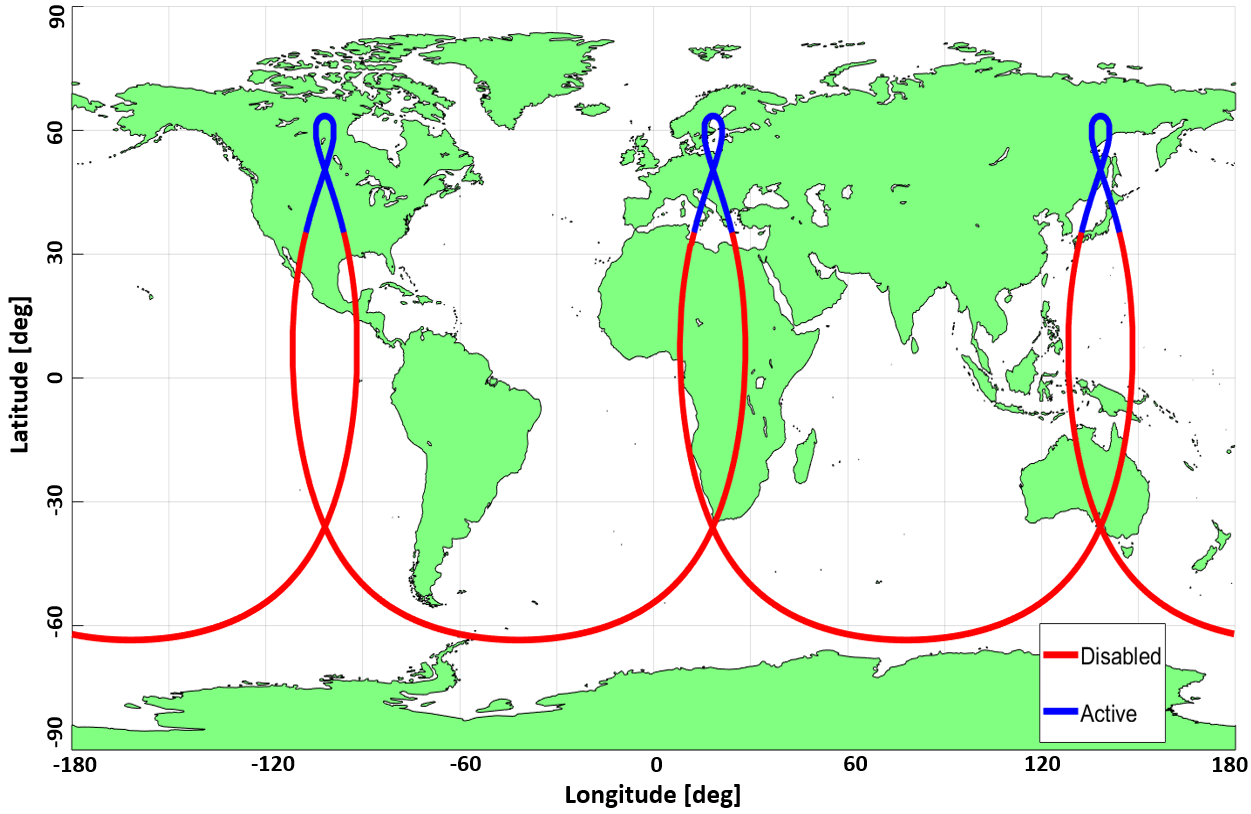}
  \caption{Ground track of one satellite from Space Norway, showing the part of the ground track when the communication payload is active (blue) and the part when the communication payload is switched off (red).}\label{fig:groundtrackheo}
\end{center}
\end{figure}

We consider eight NGSO systems which have already been approved or are waiting for approval from the FCC to operate in the Ku-band, i.e. OneWeb MEO, OneWeb LEO, SpaceX, Kepler, Theia, Karousel, Space Norway, and New Spectrum Satellite (NSS), and one GSO configuration, as illustrated in Fig.~\ref{fig:lmeoconst}. Further system parameters are summarized in the Appendix. 

The considered NGSO satellite systems cover a wide range of constellation types in terms of altitude and orbit geometry, i.e. LEO, MEO, HEO, and geosynchronous. 
Furthermore, the size of the constellations are very different, where SpaceX is the largest constellation with 4,425 satellites and Space Norway is the smallest constellation with 2 satellites. 
The considered LEO, MEO, and geosynchronous constellations are designed to cover Earth locations at latitudes between 55°S and 70°N. By contrast, the HEO constellations have different coverage areas, e.g. Space Norway intends to cover only Earth locations at latitudes above 55°N (as shown in Fig.~\ref{fig:groundtrackheo}), and NSS plans to cover Earth locations at latitudes above 42°N and above 42°S. For other locations, the communication payload of these satellites is expected to be off.  

We consider the GSO satellites Intelsat-21, Star One C1, Star One C2, Intelsat-16, SES AMC 6, Intelsat Galaxy 28, Intelsat 30, SES 1, and ANIK F1R \cite{scheds}. We assume that these satellites form a configuration with a separation angle of approximately 6° between two adjacent satellites in the equatorial plane, consistent with the regulatory requirement of a separation angle of at least 6° between two adjacent GSO satellites \cite{radioreg}. For the interested reader, results for separation angles of 10° and 20° were presented in~\cite{Braun2019}, where the interference impact was found to be lower than for 6°.

\begin{figure}[!t]
\begin{center}
\centering
  \includegraphics[width=0.9\columnwidth]{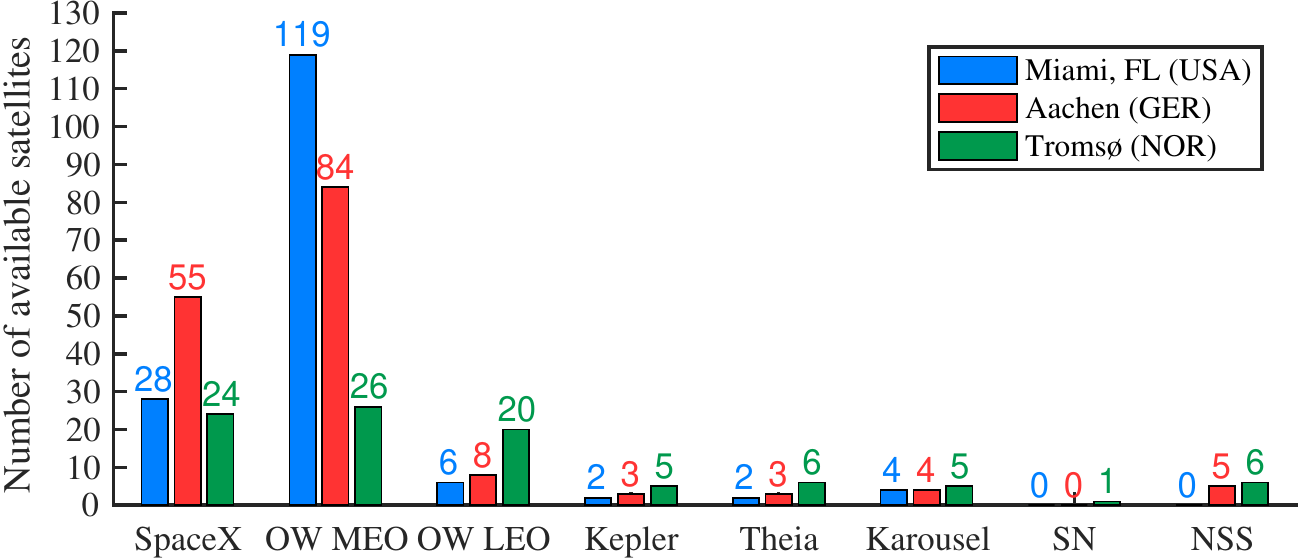}
  \caption{The average number of visible satellites for all NGSO constellations, i.e. SpaceX, OneWeb (OW), Kepler, Theia, Karousel, Space Norway (SN), and NSS, over Miami (USA), Aachen (Germany) and Troms\o~(Norway).}\label{fig:vissatcirc}
\end{center}
\end{figure}

Given the diverse covered regions on Earth of different satellite systems, we consider several locations for the ground stations, which we consider representative for studying interactions among different systems: Miami (USA), Aachen (Germany), and Troms\o~(Norway). We note that Miami is suitable for studying interactions among most NGSO systems and between NGSO-GSO, whereas Aachen and Troms\o~ are practically not affected by GSO system operation, but are relevant for HEO constellations like NSS.  
Fig.~\ref{fig:vissatcirc} shows the average number of available satellites from the NGSO systems at the three considered locations. OneWeb MEO provides the highest number of available satellites for all locations, due to the large total number of satellites, i.e. 2,560, and the higher orbital altitude than for LEO systems. SpaceX, the largest considered constellation, also provides many satellites over the three considered locations. By contrast, OneWeb LEO, Kepler, and Theia cover better locations at high latitudes, e.g. Troms\o, compared with other locations. This is due to the (nearly) polar orbits that they use. Additionally, the small constellations, i.e. Kepler and Theia, only provide a very small number of available satellites over Aachen and Miami.

\section{Simulation Setup}
\label{PerEvaMet}
We adopt the simulation approach and tool applied in~\cite{TPRC} for NGSO interference in the Ka- and V-band, extending them for the NGSO and GSO systems in the Ku-band. 
Monte Carlo simulations were conducted with the simulation tool implemented in MATLAB for satellite deployments with co-channel inter-system interference.
For all simulations, the total number of Monte Carlo iterations was set to 60,000. In the following, the simulation procedure and parameters within one Monte Carlo iteration are presented.

\label{subsec_simsteps}
\subsubsection{Selection of Satellite Positions}
For the NGSO systems, the position of the satellites in the different constellations is first selected at a random moment.
For the satellites in circular orbital planes, the satellite positions and the rotational position of the Earth are selected independently of each other, due to the short orbital period and the similar ground track for each subsequent orbital period. Consequently, a random satellite position and a random rotational position of the Earth always occur in practice after a certain amount of time. More detailed calculation steps can be found in \cite{FundofAstro}. 

This is different for the considered satellites in elliptical planes, for which only few ground tracks are possible and are repeated after one or several orbital periods. As such, the satellite position and the Earth rotation cannot be selected independently. For example, Fig.~\ref{fig:groundtrackheo} shows the ground track of one satellite from Space Norway, which is repeated after three orbital periods. Therefore, for the satellites in elliptical planes, one possible position on their orbits is randomly selected and then the corresponding rotational position of the Earth is calculated based on Kepler's equation~\cite{FundofAstro}.
The positions of the GSO satellites with respect to the Earth locations are always the same.

\subsubsection{Selection of Available Satellites}
For NGSO systems, the available satellites for a given considered ground station location are determined as the satellites that are above a preferred or at least above a minimum elevation angle from the point of view of the ground station. These angles are specified by each NGSO operator in~\cite{scheds}, where the minimum elevation angles range from around 10° for small NGSO constellations to 40°--55° for large constellations. The available satellites are determined by calculating the slant range between the considered ground station location and the satellite position~\cite{slrangecalc}.
We assume that each GSO satellite is active and has one associated ground station at a considered Earth location.
\subsubsection{Link Selection}\label{linksel}
For each victim NGSO constellation, one satellite from the victim constellation and one satellite from each interfering NGSO constellation are selected out of those available. 
For the victim constellation, one random available satellite is selected when no interference mitigation technique is applied. For the look-aside mitigation technique, one random available satellite out of those which have a separation angle of at least 5° to all interfering satellites is selected, if possible.\footnote{If no victim satellite fulfils this requirement, the available satellite with the separation angle closest to 5° is selected.} When the GSO protection technique is used, one random available satellite out of those which have a separation angle of 30° to all GSO satellites is selected. We note that for band-splitting, the link selection is similar to the case where no interference mitigation technique is applied.  

Since it is not clear how many NGSO satellites will actually be active in practice, only three satellites for each interfering constellation are considered as active interferers for a given victim satellite system, as found reasonable in~\cite{TPRC}. These three satellites are randomly selected from those available. Furthermore, we assume that only one of these three actively interfering satellites forms a link with a ground station that is co-located with the ground station of the victim system~\cite{TPRC}. 
To this end, the actively interfering satellite that causes the strongest interference to the selected communication link of the victim satellite system is selected. Therefore, the closest actively interfering satellite to the victim satellite in terms of angular separation is selected\footnote{Thus, we assume that the other two actively interfering satellites from each interfering constellation do not strongly interfere the victim system~\cite{TPRC}.}~\cite{slrangecalc}. 
For the GSO configuration we assume that all GSO satellites communicate simultaneously with their corresponding ground stations. 
\subsubsection{Link Budget Analysis}\label{linkbudpara}
The received power (useful or interference) at the ground stations in the downlink is
\begin{equation}
P_{GS}=fcn(EIRPD_{S}, L_{FS}, A_T, G_{GS}, T_{GS}, \psi_{GS}),
\end{equation}
where $EIRPD_{S}$ is the effective isotropically radiated power density of the satellite, $L_{FS}$ is the free-space path loss~\cite{SatComSysEng}, $A_T$ is the atmospheric attenuation, $G_{GS}$ is the gain of the receive antenna at the ground station, $T_{GS}$ is the ground station receiver noise temperature, and $\psi_{GS}$ is the 3~dB beamwidth of the ground station receiver. 
We set the downlink frequency to $f$=12~GHz~\cite{TPRC} and we model $A_{T}$ with the ``ITU-R Propagation Models Software Library''~\cite{CNESSoft}, which takes into account rain, cloud, gas, and scintillation models. Table~\ref{atmattloc} summarizes examples of atmospheric attenuation for the considered ground station locations, where Miami has the poorest atmospheric conditions due to its specific climate. For each link we consider a random and uniformly distributed unavailability probability in the interval (0,1). 

\begin{table}[!t]
\caption{Example of atmospheric attenuation for different Earth locations, for a frequency of 12~GHz, a satellite elevation angle of 50°, an antenna diameter of 0.45~m, and an unavailability probability of the communication link of 0.1\%.}
\begin{center}
\begin{tabular}{|c|c|c|c|}
\hline
&&&\textbf{Atmospheric} \\
\textbf{Location} & \textbf{Latitude}& \textbf{Longitude} & \textbf{attenuation} \\
\hline
Troms\o~(Norway)			&  69.7°N	&     18.9°E  & 0.9~dB   \\
\hline          
Aachen (Germany)			&  50.8°N	&     6.1°E  & 1.7~dB   \\ 
\hline
Miami, FL (USA)				&  26.8°N	&     80.2°W  & 4.8~dB   \\      
\hline
\end{tabular}
\label{atmattloc}
\end{center}
\end{table}

The required satellite transceiver parameters for different operators are available in~\cite{scheds}, for the transmitter ($EIRPD_{S}$, antenna gain, EIRP) and receiver (antenna gain, G/T, saturation flux density). By contrast, in~\cite{scheds} there is less information about their ground stations (only the antenna diameter), except for SpaceX (specifying the antenna gains and EIRPD), which has already applied for ground station licences. As such, for the ground stations of other operators we calculate the antenna gain $G_{GS}$ of the user terminals as
\begin{equation}
G_{GS} = A_{Eff}\times \bigg(\frac{\pi A_{dia} f}{c}\bigg)^2~,
\label{eq:gaingsantenna}
\end{equation} 
where $A_{Eff}$ is the antenna efficiency set to 80\%~\cite{TPRC}, $A_{dia}$ is the antenna diameter, and $c$ is the speed of light~\cite{SatComSysEng}. Further, a noise temperature of $T_{GS}$=140~K is assumed~\cite{IntroSat}. We note that we consider only small user terminals as ground stations of NGSO systems, since more user terminals than Earth stations are expected in practice, so the risk of interference is higher for user terminals. 
The considered antenna patterns for all the NGSO and GSO satellite and ground station antennas (including the 3 dB beamwidth) are based on ITU-R recommendations in~\cite{itu1528}. For the GSO satellite system, we consider dish antennas for the ground stations, where we analyse both user terminals (diameter: 0.75~m) and Earth stations (diameter: 3.7~m). 
The most important satellite and ground station transceiver parameters are summarized in the Appendix. 

For the NGSO and GSO systems we first consider the original transceiver parameters proposed in~\cite{scheds}. 
For the satellite transceivers of NGSO constellations in circular LEO and MEO planes and for all the NGSO ground station transceivers, we also consider a tuned set of parameters aiming at harmonization among systems based on their altitudes, \emph{cf.} Appendix.

\section{Results}\label{results}
This section presents and discusses a representative selection of our performance evaluation results for co-channel, co-polarized interference among the considered NGSO-NGSO and NGSO-GSO satellite systems. Extended results were presented in~\cite{Braun2019}. We quantify the interference impact in terms of throughput degradation (\emph{cf.} Section~\ref{EvalMet}), where we present results as a complementary cumulative distribution function (CCDF) for the 60,000 Monte Carlo iterations. 
We note that this representation is consistent with the charts for \emph{risk-informed interference assessment} in~\cite{fccrisk}. 
In the following, simulation results for the original and tuned transceiver parameters are presented in Section~\ref{resorig} and~\ref{resadap}, respectively.

\subsection{Original NGSO Transceiver Parameters}\label{resorig}

\begin{figure}[t]
\begin{center}
\centering
  \subfloat{\includegraphics[width=0.95\columnwidth]{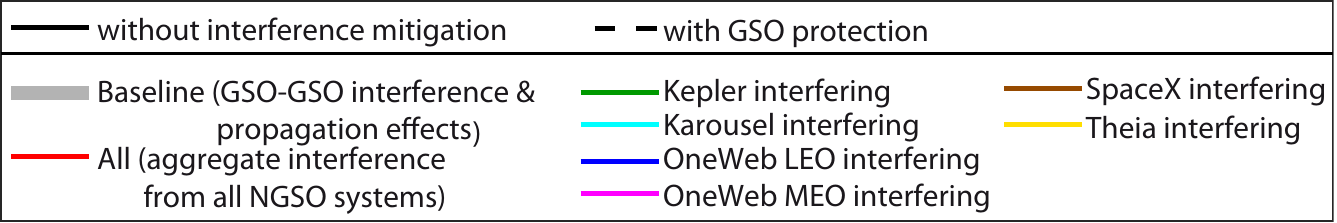}}
  \\
  \subfloat{\includegraphics[width=0.95\columnwidth]{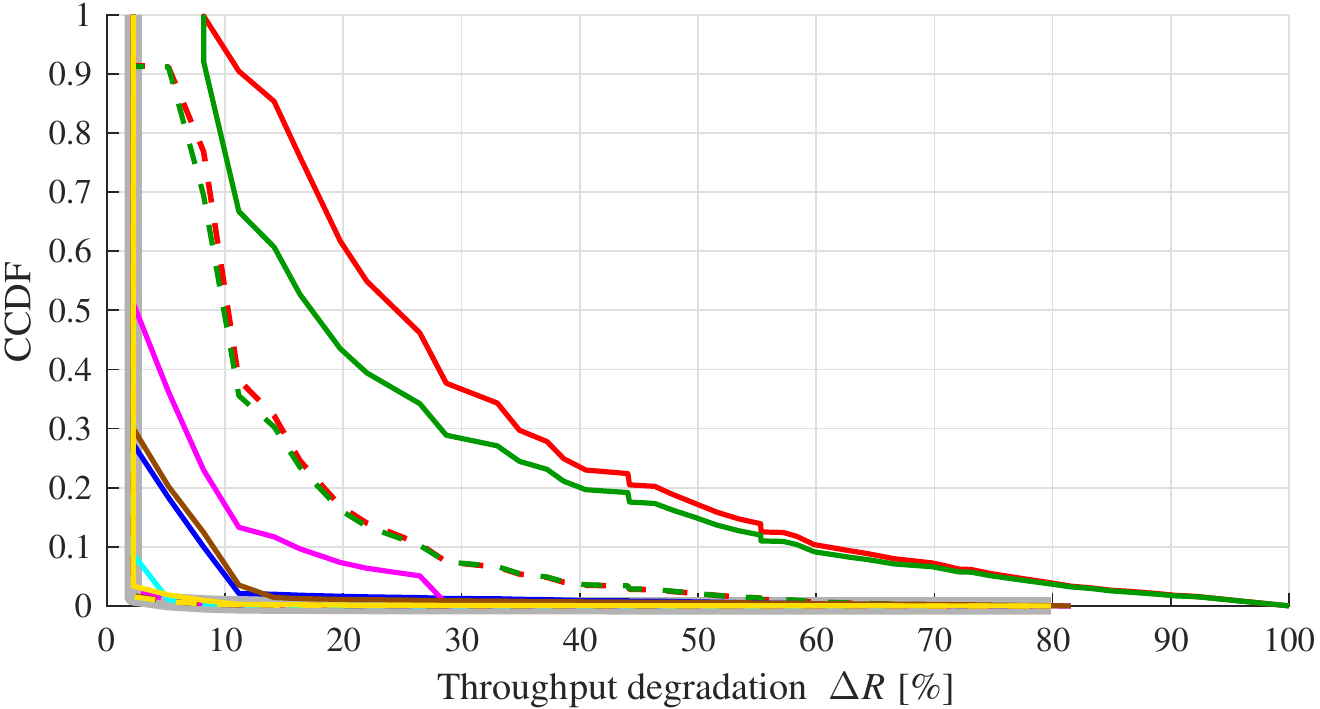}}
  \caption{Distribution of throughput degradation at the GSO ground station of Intelsat-16 in Miami, where all NGSO systems that cover this location are interferers in the downlink. Results are shown for the GSO baseline, and for all or pairs of GSO-NGSO coexisting systems, without interference mitigation and for the GSO protection technique. The GSO ground station is a user terminal and the NGSO systems apply the \textbf{original transceiver parameters}.}\label{fig:gsongsoorigint16}
\end{center}
\end{figure}

We present NGSO-GSO and NGSO-NGSO interference simulation results generated for the original NGSO transceiver parameters as proposed by the satellite operators in~\cite{scheds} and summarized in the Appendix. 
We first focus on the scenario where NGSO and GSO systems coexist. Fig.~\ref{fig:gsongsoorigint16} shows the distribution of the throughput degradation for a GSO user terminal belonging to Intelsat-16, for the GSO configuration coexisting with all considered NGSO constellations in the downlink in Miami.\footnote{We note that Space Norway and NSS are expected to be turned off above Miami, so they are not considered for this location.} We note that Miami is selected as a representative example of a location that is covered by GSO satellite systems. This figure shows the results for coexistence among all and between pairs of GSO-NGSO satellite systems, for the case of no interference mitigation and for the GSO protection technique. Furthermore, results for the baseline scenario of the standalone GSO system, i.e. only GSO-GSO interference, are also shown.

The throughput degradation for the baseline is in most cases very low (i.e. the throughput degradation exceeds 2\% in only 1\% of the cases), which shows that the impact of GSO interference and propagation and atmospheric attenuation on the GSO system is negligible.
The highest throughput degradation is observed when the GSO system coexists with all NGSO systems that do not apply any interference mitigation technique, i.e. a median throughput degradation of 27\%. This is expected since the level of interference at the GSO ground station is higher. 
For coexistence between pairs of GSO-NGSO systems, the median throughput degradation is lower (even without interference mitigation), i.e. between 2\% and 20\%, where Kepler causes the highest degradation that is only 7~percentage points (pp) lower than for the case where all NGSO systems coexist.
This is due to the originally proposed transceiver parameters for Kepler (\emph{cf.} Table~\ref{tab:origngsosat}), where the satellites of Kepler have the highest EIRPD of all LEO and MEO systems (i.e. -21.4~dBW/Hz), although they are LEO satellites operating at a low altitude of 600~km. 
Importantly, this shows that Kepler is the dominant interfering system. 
Furthermore, when the NGSO systems apply the GSO protection technique, the median throughput degradation decreases to 11\% when all NGSO systems are active and is between 2--11\% for coexistence with a single NGSO system. We emphasize that, for this case, Kepler causes a throughput degradation as high as that caused by all NGSO systems.
This is an important result and shows that although the GSO protection technique reduces the interference from NGSO at the GSO system in the downlink, Kepler still causes a rather high throughput degradation, whereas interference from all other NGSO systems is negligible. 
This suggests that in practice, the current spectrum regulation to protect GSO systems in the Ku-band may be insufficient.  
We note however, that for GSO configurations with a larger angular separation, i.e. 10° and 20°, or for Earth stations (instead of user terminals) a lower throughput degradation was observed~\cite{Braun2019}. These results are not shown here for brevity.

\begin{figure}[t]
\begin{center}
\centering
  \subfloat{\includegraphics[width=0.85\columnwidth]{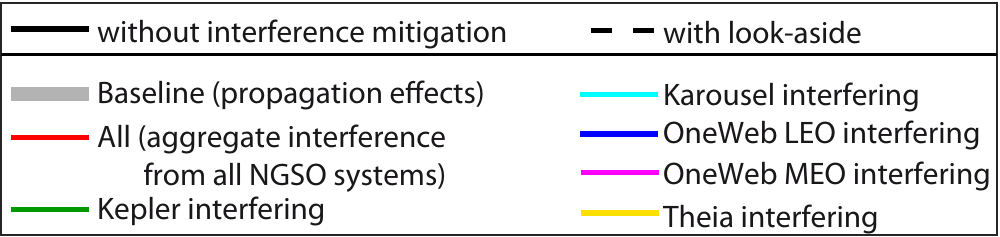}}
  \\
  \subfloat{\includegraphics[width=0.95\columnwidth]{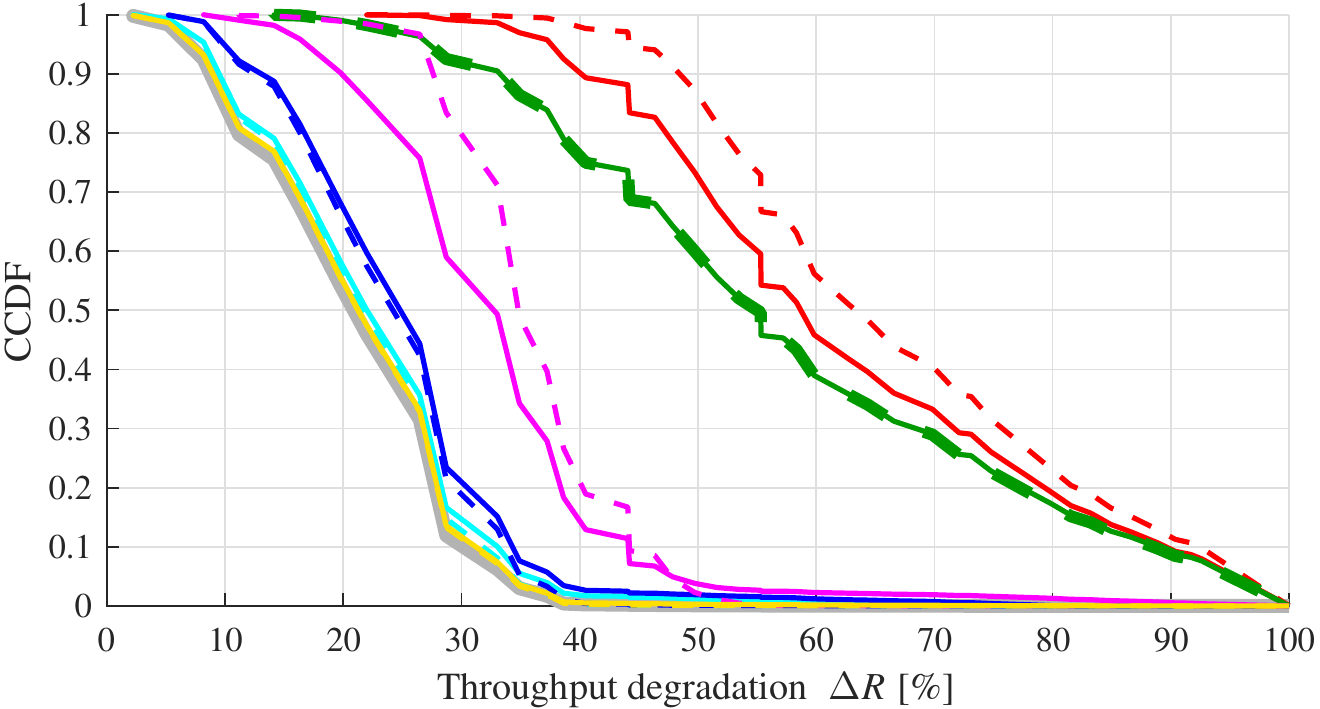}} 
  \caption{Distribution of throughput degradation at the ground station of SpaceX in Miami, where all other NGSO systems that cover this location are interferers in the downlink. Results are shown for the SpaceX baseline, and for all or pairs of NGSO-NGSO coexisting systems, without interference mitigation and for look-aside. The NGSO systems apply the \textbf{original transceiver parameters}.}\label{fig:spacexorigdl}
\end{center}
\end{figure}

Let us now consider the impact of NGSO-NGSO interference.
Fig.~\ref{fig:spacexorigdl} shows the distribution of the throughput degradation at a ground station of SpaceX in Miami when all NGSO constellations coexist (SpaceX, OneWeb LEO, OneWeb MEO, Kepler, Theia, and Karousel) without any interference mitigation technique, or with look-aside. 
The throughput degradation for the baseline scenario is rather high, e.g. the median degradation is 22\%, which shows that the impact of propagation and atmospheric attenuation on SpaceX is much stronger than on the GSO system in Fig.~\ref{fig:gsongsoorigint16}, due to the lower $EIRPD_{S}$ used by SpaceX.
Furthermore, when coexisting with other NGSO systems, the ground station of SpaceX is strongly interfered with and predominantly by Kepler, e.g. a median throughput degradation of 55\% when coexisting only with Kepler compared with 60\% when coexisting with all constellations, without interference mitigation. 
Moreover, the throughput degradation is not decreased by look-aside when SpaceX coexists only with Kepler. 
This is consistent with the results for the GSO ground station in Fig.~\ref{fig:gsongsoorigint16} and confirms that Kepler is a strong interferer also for NGSO-NGSO coexistence, due to its satellite transmitter parameters. 

Furthermore, the throughput degradation of SpaceX when coexisting with all NGSO constellations for the look-aside mitigation technique is slightly larger than when interference is not mitigated, e.g. 65\% median throughput degradation with look-aside vs. 60\% without interference mitigation. 
By comparing the results when SpaceX coexists with only one other NGSO system in Fig.~\ref{fig:spacexorigdl}, we observe a similar trend (of larger degradation for look-aside vs. no interference mitigation) only for coexistence with OneWeb MEO, e.g. a median degradation of 37\% for look-aside and of 34\% without interference mitigation. 
This is due to the fact that the ground station of SpaceX selects a satellite at a quite low elevation angle, i.e. on average 48.6°, when applying look-aside and coexisting with OneWeb MEO, since OneWeb MEO is a large constellation and SpaceX needs to select a satellite that is 5° away from all OneWeb MEO satellites. By contrast, SpaceX uses a satellite at an average elevation angle of 58° without interference mitigation, which results in a lower path loss than for the lower elevation angle with look-aside. 
These results show overall that using look-aside does not always improve the overall satellite throughput performance, especially for cases where the path loss has a dominant effect over NGSO-NGSO interference.       

The results in Figs.~\ref{fig:gsongsoorigint16} and \ref{fig:spacexorigdl} show overall that Kepler is the strongest interfering constellation in the downlink and causes significant interference to both GSO systems and other NGSO systems, even when interference mitigation techniques based on satellite diversity are applied. Since this is chiefly due to the satellite transmitter parameters that were originally proposed for Kepler in~\cite{scheds}, in the following section we present results for tuned NGSO transceiver parameters, which enable us to harmonize the impact of the transceiver parameters $EIRPD_S$ and $G$. We can thus study in more detail the impact of the constellation design in terms of number of satellites, satellite elevation angles, or orbit geometry.\footnote{We note however, that for NGSO constellations in elliptical planes, i.e. Theia, Space Norway, NSS, and Karousel, we always apply the original satellite transceiver parameters, since these constellations do not operate at a fixed altitude and would thus require e.g. power control for parameter harmonisation based on the altitude.}
Moreover, this also enables us to investigate whether the GSO system could indeed be protected (as the current FCC regulation requires) via the simple satellite diversity technique with a minimum separation angle between NGO-GSO of 30°, as proposed by different operators in~\cite{scheds}.

\subsection{Tuned NGSO Transceiver Parameters}\label{resadap}

\begin{figure}[t]
\begin{center}
\centering
  \subfloat{\includegraphics[width=0.9\columnwidth]{gsovsallorigdl_lgd}}
  \\
   \subfloat{\includegraphics[width=0.95\columnwidth]{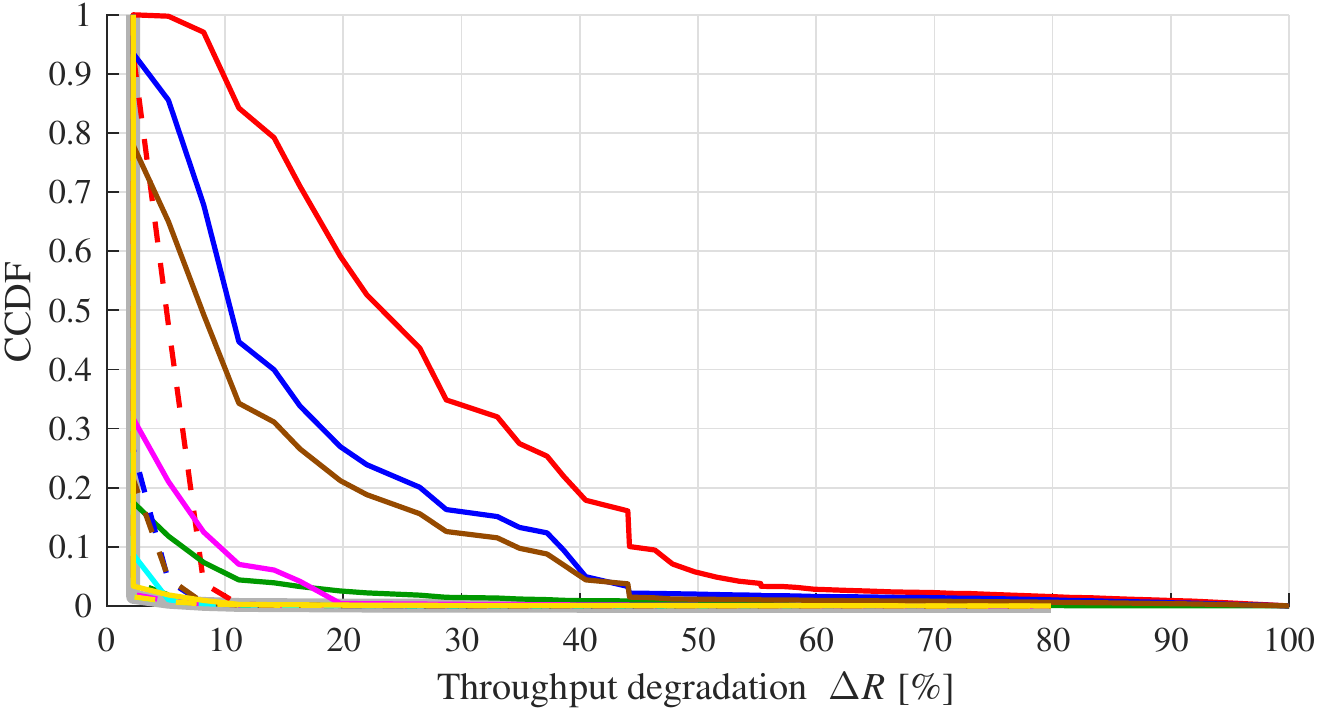} }
  \caption{Distribution of throughput degradation at the GSO ground station of Intelsat-16 in Miami, where all NGSO systems that cover this location are interferers in the downlink. Results are shown for the GSO baseline, and for all or pairs of GSO-NGSO coexisting systems, without interference mitigation and for the GSO protection technique. The GSO ground station is a user terminal and the NGSO systems apply the \textbf{tuned transceiver parameters}.}\label{fig:gsovsalladapgdl}
\end{center}
\end{figure}

We present and discuss the simulation results for tuned NGSO transceiver parameters summarized in the Appendix. 
Let us first consider the GSO system.
Fig. \ref{fig:gsovsalladapgdl} shows the distribution of the throughput degradation at a GSO ground station in Miami, where all NGSO systems are interferers and apply the tuned transceiver parameters. We note that these results correspond to those in Fig.~\ref{fig:gsongsoorigint16}, where the original NGSO transceiver parameters were applied. 
The highest throughput degradation is observed when all systems coexist without any interference mitigation technique, e.g. a median degradation of 28\%, due to the high aggregate level of interference, as expected. 
By comparing the results for pairs of coexisting GSO-NGSO satellite systems, we observe that SpaceX and OneWeb LEO are the main interfering systems resulting in a median degradation of 8\% and 10\%, respectively. 
However, when the GSO protection technique is applied, the median throughput degradation is reduced to 5\% when all considered NGSO satellite systems cause interference to the GSO ground station, and the degradation exceeds 10\% only for a negligible number of cases. 
This shows that the GSO protection technique is more efficient in the downlink when the NGSO systems apply the tuned set of transceiver parameters compared to the original transceiver parameters (\emph{cf.} Fig.~\ref{fig:gsongsoorigint16}), but interference from NGSO systems is not completely mitigated at the GSO user terminal.

\begin{figure}[t]
\begin{center}
\centering
\subfloat[SpaceX]{\includegraphics [width=0.95\columnwidth] {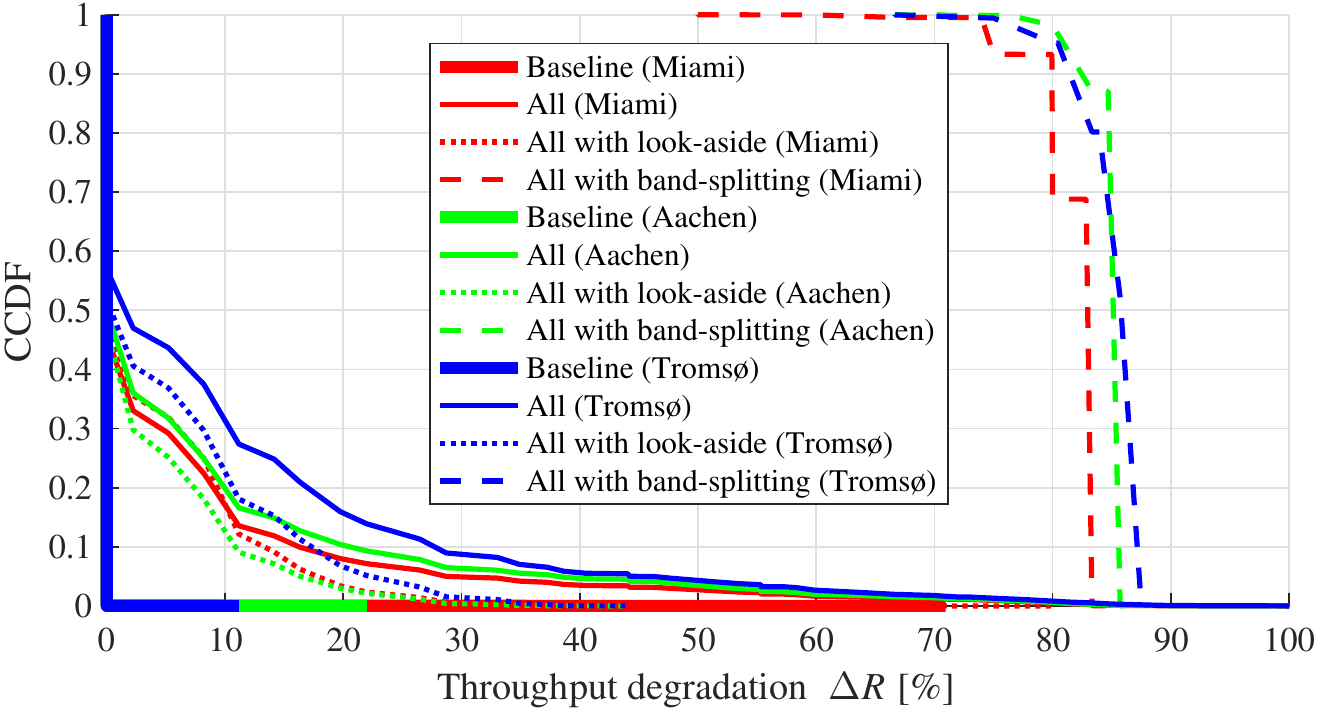}\label{subfig:spacex}}
\\
\subfloat[Kepler]{\includegraphics [width=0.95\columnwidth] {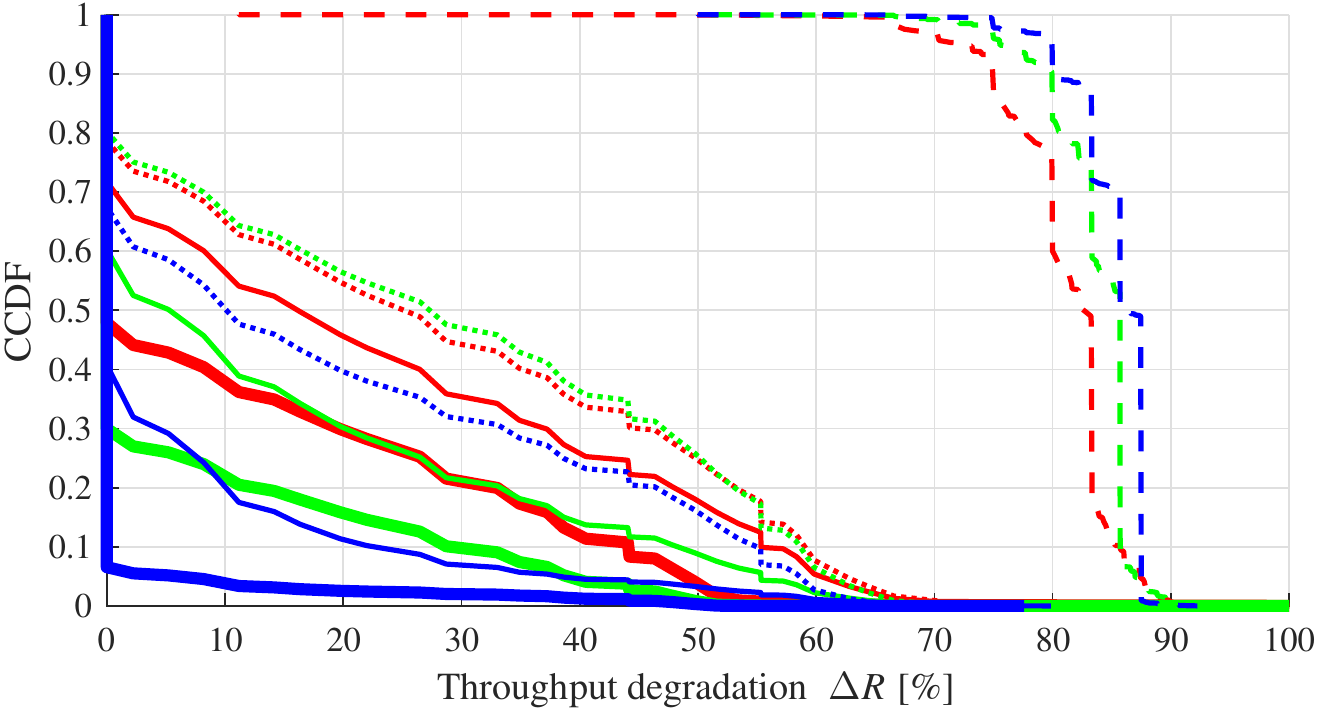}\label{subfig:kepler}}
\\
\subfloat[NSS]{\includegraphics [width=0.95\columnwidth] {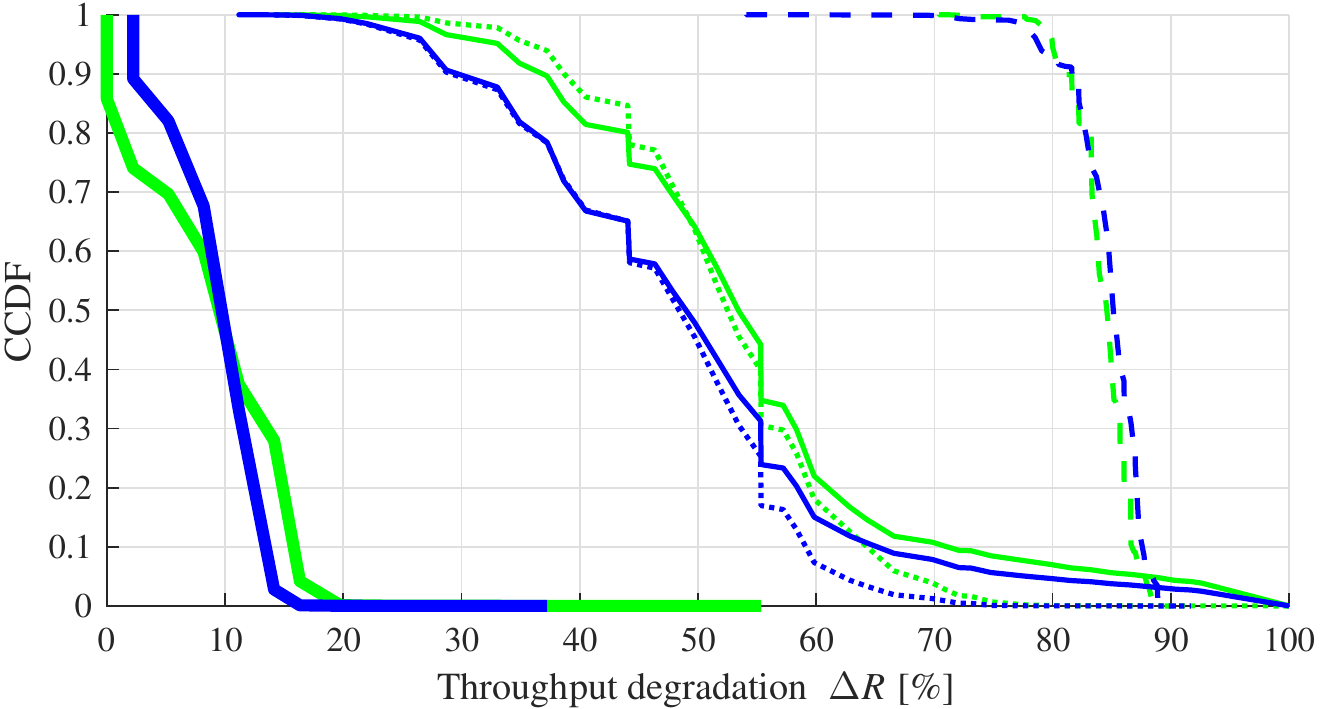}\label{subfig:nss}}
\caption{Throughput degradation at the ground stations of SpaceX, Kepler, and NSS in Miami, Aachen and Troms\o, for \textbf{tuned transceiver parameters}. All other NGSO satellite systems covering these locations are interferers. Results are shown for the baseline, and for all NGSO-NGSO coexisting systems, without interference mitigation and for look-aside, or band-splitting.}\label{fig:spacexdifflocadap}
\end{center}
\end{figure}

Let us now focus on the impact of NGSO-NGSO interference for different satellite constellations.
Fig.~\ref{fig:spacexdifflocadap} shows the distribution of the throughput degradation at ground stations from SpaceX, Kepler, and NSS, when all NGSO constellations coexist in the downlink in Miami, Aachen, and Troms\o.\footnote{We note that NSS does not cover Miami.} 
We first consider the throughput degradation at the ground station of SpaceX in Fig.~\ref{subfig:spacex}.
The throughput degradation for the baseline scenario is reduced to nearly zero for all locations, in contrast to the original transceiver parameters as shown in Fig.~\ref{fig:spacexorigdl}.
This shows that carefully selecting the transceiver parameters can compensate for the effects of path loss and atmospheric attenuation.
The throughput degradation in Fig.~\ref{subfig:spacex} is rather low when all systems coexist without interference mitigation, regardless of the location, i.e. the highest median degradation is 2\% in Troms\o.
This result is consistent with Fig.~\ref{fig:vissatcirc}, which shows that SpaceX covers Troms\o~with fewer satellites than Aachen and Miami.
Further, the look-aside mitigation technique benefits SpaceX especially for the higher range of throughput degradation, e.g. in Troms\o~the maximum throughput degradation is at most 45\% with look-aside compared with 100\% without interference mitigation. 
We note that look-aside benefits SpaceX for all considered locations including Miami, unlike for the original transceiver parameters in Fig.~\ref{fig:spacexorigdl}. 
The results for SpaceX with band-splitting in Fig.~\ref{subfig:spacex} show a significant increase in throughput degradation, i.e. a median throughput degradation of at least 83\% for all locations, which suggests that band-splitting will not be preferred in practice. Importantly, these results are consistent for Kepler and NSS in Figs.~\ref{subfig:kepler} and~\ref{subfig:nss}, respectively, and confirm the results for the Ka- and V-band in~\cite{TPRC}.
As an insight, we observe that band-splitting is triggered for all cases (i.e. the degradation with band-splitting is always above 50\%), although SpaceX does not always suffer from significant interference (i.e. the throughput degradation is zero for at least 45\% of the cases without interference mitigation). This is due to the fact that band-splitting is also triggered when SpaceX causes strong interference to another system and shows that SpaceX is a strong interferer for other NGSO systems.

Fig. \ref{subfig:kepler} shows the distribution of the throughput degradation at the ground station of Kepler. 
Unlike for SpaceX, the baseline throughput of Kepler varies significantly for the considered locations: the highest degradation occurs in Miami (degradation larger than zero for 49\% of the cases), a moderate degradation is observed in Aachen (degradation larger than zero for 30\% of the cases), and the lowest degradation is observed in Troms\o~(degradation larger than zero for 8\% of the cases).
This is a result of the constellation geometry. Specifically, Kepler has rather few satellites and with polar orbits, which benefit the locations at higher latitudes like Troms\o. Consequently, Kepler uses satellites with different average elevation angles of 44° and 30° in Troms\o~and Miami, respectively, where smaller angles result in a higher propagation attenuation.  
These results are consistent with the average number of available satellites in Miami, Aachen, and Troms\o~(\emph{cf.} Fig.~\ref{fig:vissatcirc}).  
For coexistence with the other NGSO systems without interference mitigation, we observe the same trend with respect to location: the highest throughput degradation occurs in Miami (e.g. a median degradation of 17\%), and the lowest in Troms\o~(a median degradation of zero). 
For look-aside, the throughput degradation for Kepler is significantly increased compared with no interference mitigation, regardless  of the location, e.g. in Aachen the median degradation is 28\% with look-aside and 5\% without interference mitigation. This occurs due to the small number of available satellites, so the ground station of Kepler must select a satellite with a very low elevation angle when the look-aside mitigation technique is applied. 
These results show overall that the throughput performance of satellite systems and of look-aside are very sensitive to the joint effect of the number of satellites, the geometric properties of the orbits, and the location of the ground stations.   

Fig.~\ref{subfig:nss} shows the distribution of the throughput degradation for the ground station of NSS. 
The baseline results are similar for the two considered locations that are covered by NSS, i.e. Aachen and Troms\o, with a median throughput degradation of 10\%. We note that for NSS, the communication link is most affected when a satellite at a high altitude is selected, where the altitude varies between 17,200~km and 26,700~km for NSS, due to its elliptical orbits. Despite this large variation which affects the path loss, the same transceiver parameters are specified in~\cite{scheds}. Power control could be a solution to decrease the baseline throughput degradation in this case.
Importantly, without interference mitigation, NSS is strongly interfered by the other NGSO systems, i.e. the median throughput degradation is 54\% and 49\% in Aachen and Troms\o, respectively. Look-aside decreases the throughput degradation of NSS only marginally, and typically for the range of large throughput degradation (i.e. above 50\%). This is due to the low number of available satellites, i.e. on average 5 and 6 available satellites over Aachen and Troms\o, respectively. These results confirm that look-aside is not efficient at mitigating interference for constellations with a small number of satellites, so more sophisticated interference mitigation techniques are required for such cases.

\section{Conclusions \& Discussion}\label{conclusions}
We presented an extensive inter-satellite interference study for new NGSO systems coexisting in the Ku-band in the downlink, where we also considered some existing GSO systems. 
Our results showed that current spectrum regulation is not always sufficient to ensure GSO protection from NGSO interference, especially due to the large specified transmit power of the already approved LEO Kepler constellation. This suggests that the impact of inter-satellite interference is very sensitive to the joint effect of the transceiver parameters and the orbit type.  
Furthermore, this may also result in strong downlink interference towards other NGSO systems, where simple interference mitigation techniques like look-aside are not sufficient.
This opens the question of whether spectrum regulators should impose restrictions on the satellite transmit power based on the altitude and orbit type.

For tuned NGSO transceiver parameters (i.e. aiming at better harmonization among constellations at the same altitude), the efficiency of the simple look-aside technique is strongly coupled with the size and geometry of an NGSO system, and with the location of the ground stations. For large systems like SpaceX, look-aside reduces the throughput degradation due to interference, regardless of the ground location, whereas for small constellations with polar or elliptical orbits like Kepler and NSS, it is sometimes even preferable to suffer from interference than to apply look-aside. 
Furthermore, we confirm the findings in~\cite{TPRC} for the Ka- and V-band that band-splitting among satellite operators significantly degrades throughput, also for the Ku-band.  
Our results show overall that there are reasons to worry about interference in emerging dense NGSO constellations, since the complexity of the inter-satellite interactions for these new deployments is too high to be managed via the simple interference mitigation techniques that have largely been applied so far. Consequently, more sophisticated engineering solutions and potentially stricter regulatory requirements are needed.

\section*{Acknowledgement}
We thank Susan Tonkin and Pierre de Vries for useful discussions and for their publicly available simulator in Matlab.

\section*{Appendix}
\label{sec:appendix}
Tables~\ref{satorbconstcirc}--\ref{tab:origngsogs} summarize the satellite constellation and transceiver parameters.
For the downlink, the satellite transmitter and ground station receiver parameters in Tables~\ref{tab:origngsosat}-\ref{tab:origngsogs} are relevant.
The original transceiver parameters are based on~\cite{scheds, itu1528}. 
The tuned transceiver parameters of the NGSO satellites were selected as reasonable values based on the original parameters and such that there are large antenna gains rather than a high transmit power~\cite{ITU-R1999}, aiming for harmonization across systems at different altitudes. For instance, OneWeb LEO and SpaceX use the same tuned transceiver parameters, since their satellites are at similar altitudes.
The tuned transceiver parameters of the NGSO ground stations are selected with respect to the original parameters of SpaceX, which is the only operator already specifying these parameters in~\cite{scheds}, and such that harmonization is achieved across systems at different altitudes. 
The transceiver parameters for the GSO user terminals and Earth stations are always based on the real ground stations in the Ku-band in \cite{antgsout} and \cite{antgsoes}, respectively.

\begin{table}[!t]
\caption{Properties of constellations in circular and elliptical orbital planes for the considered NGSO systems}
\begin{center}
\begin{tabular}{|p{0.2cm}|c|c|c|c|c|}
\cline{2-6}
 \multicolumn{1}{c|}{}
 & \textbf{System}
 & \textbf{Orbit} 
 & \parbox[c][0.8cm][c]{0.8cm}{\centering \textbf{No. of}\\ \textbf{satellites}}
 &  \parbox[c][0.8cm][c]{1cm}{\centering \textbf{Altitude} \\ \textbf{[km]}} 
 & \parbox[c][0.8cm][c]{1cm}{\centering \textbf{FCC} \\ \textbf{approval}} \tabularnewline
\hline
 \multirow{4}{*}{\rotatebox[origin=c]{90}{\textbf{circular}}} 
 & OneWeb 		 	& 	MEO			&  		2,560	& 8,500 & pending\\
\cline{2-6}
& OneWeb 		 	& 	LEO			&  		1,980  	& 1,200 & partial\\
\cline{2-6}
& SpaceX	 		& 	LEO			&  		4,425	& 1,200 &  approved\\
\cline{2-6}
& Kepler 		 	& 	LEO		&  		140		& 600 & approved\\            
\hline
\multirow{6}{*}{\rotatebox[origin=c]{90}{\textbf{elliptical}}}
& Theia 		 	& 	LEO	&  		112		& \parbox{1.8cm}{apogee: 809 \\ perigee: 791} & approved \\
\cline{2-6}
& Karousel  		&	\parbox{1.3cm}{Geo-synchronous}   &  		12  	& \parbox{1.8cm}{apogee: 40,002 \\ perigee: 31,569} & approved \\
\cline{2-6}
& SN 	& 	HEO		&  		2	 & \parbox{1.8cm}{apogee: 43,509  \\  perigee: 8,089} & approved\\
\cline{2-6}
& NSS 		& 	HEO		&  		15		& \parbox{1.8cm}{apogee: 26,190 \\ perigee: 1,650} & pending\\          
\hline
\end{tabular}
\label{satorbconstcirc}
\end{center}
\end{table}

\begin{table}[!t]
\caption{Original (green) and tuned (red) transceiver parameters for NGSO satellites and original parameters for GSO satellites (yellow), where $EIRPD$ is the max. effective isotropically radiated power density, and $G$ is the max. antenna gain }
\begin{center}
\begin{tabular}{|c|c|c|c|c|}
\hline 
 &\multicolumn{2}{|c|}{\textbf{Satellite Tx}}&\multicolumn{2}{|c|}{\textbf{Satellite Rx}} \\
\cline{2-5} 
 \textbf{System} 
 & \parbox[c][0.8cm][c]{0.5cm}{\centering $G$ \\ (dBi)} 
 & \parbox[c][0.8cm][c]{1.5cm}{\centering $EIRPD$ \\(dBW/Hz)}  
 & \parbox[c][0.8cm][c]{0.5cm}{\centering $G$ \\ (dBi)} 
 & \parbox[c][0.8cm][c]{0.8cm}{\centering $(G/T)$ \\ (dB/K) } \\
\hline
 {\cellcolor{green!25}} SpaceX		 	& 		37.1	&  		-47.1  & 		37.1	&  		9.8	\\
\hline
 {\cellcolor{green!25}} OneWeb (OW) MEO		 	& 		49.1	&  		-25.6 & 		50.6	&  		23.7	\\
\hline
 {\cellcolor{green!25}} OneWeb (OW) LEO		 	& 		24.5	&  		-49.4	& 		26.0	&  		-1.0  \\
\hline
 {\cellcolor{green!25}} Kepler		 	& 		23.6	&  	-21.4 & 		25.5	&  		-6.3		\\
\hline
 {\cellcolor{green!25}} Theia		 	& 		34.3	&  		-52.5  & 		28.4	&  		1.75	\\
\hline
 {\cellcolor{green!25}} Space Norway (SN)	& 		35.0	&  	-27.6 & 		35.0	&  		8.0	 \\
\hline
 {\cellcolor{green!25}} NSS		 	& 		41.0	&  		-21.6 & 		42.5	&  		14.7	   \\ 
\hline
 {\cellcolor{green!25}} Karousel 	& 		34.1	&  		-24.6	& 		38.3	&  		11.8   \\ 
\hline
\hline
 {\cellcolor{red!25}} SpaceX 		 	& 		50	&  		-38  & 		40	&  		10	\\
\hline
 {\cellcolor{red!25}} OneWeb MEO		 	& 		55	&  		-29 & 		45	&  		15	\\
\hline
 {\cellcolor{red!25}} OneWeb LEO		 	& 		50	&  		-38	& 		40	&  		10  \\
\hline
 {\cellcolor{red!25}} Kepler		 	& 		50	&  		-40 & 		40	&  		10		\\
\hline
\hline
 {\cellcolor{yellow!25}} Intelsat-16 (long. 76.2~W)		 	& 		32	&  		-16.9  & 	35	&  		11.5  \\
\hline
\end{tabular}
\label{tab:origngsosat}
\end{center}
\end{table}

\begin{table}[!t]
\caption{Original (green) and tuned (red) transceiver parameters for NGSO ground stations, and selected parameters for GSO ground stations, where $\psi$ is the 3~dB beamwidth}
\begin{center}
\begin{tabular}{|p{1.2cm}|c|p{0.4cm}|c|c|c|p{0.7cm}|c|}
\hline
 & \textbf{Ant.} &\multicolumn{3}{|c|}{\textbf{Ground station Tx}}&\multicolumn{3}{|c|}{\textbf{Ground station Rx}} \\
\cline{3-8} 
 \centering \textbf{System} 
 & \parbox[c][0.8cm][c]{0.5cm}{\centering \textbf{diam.} \\ (m)}  
 & \parbox[c][0.8cm][c]{0.5cm}{\centering $G$ \\ (dBi)}  
 & \parbox[c][0.8cm][c]{1.3cm}{\centering $EIRPD$ \\(dBW/Hz)} 
 & \parbox[c][0.8cm][c]{0.3cm}{\centering $\psi$ \\ (°) } 
 & \parbox[c][0.8cm][c]{0.5cm}{\centering $G$ \\ (dBi)} 
 & \parbox[c][0.8cm][c]{0.7cm}{\centering $G/T$ \\ (dB/K) } 
 & \parbox[c][0.8cm][c]{0.3cm}{\centering $\psi$ \\ (°) } \\
\hline
 {\cellcolor{green!25}} SpaceX & 0.45  & 35.4	& 	-40 & 1.7  		& 	34	&  		12.6 & 1.9 \\
\hline
 {\cellcolor{green!25}} OW MEO & 0.45  & 35.4	& 	-33.4//-23.4 & 1.7  & 	34	&  		12.6 & 1.9 \\	
\hline
 {\cellcolor{green!25}} OW LEO & 0.45  & 35.4	& 	-17.9//-17.8 & 1.7  & 	34	&  		12.6 & 1.9 \\
\hline
 {\cellcolor{green!25}} Kepler & 0.3  & 31.2	& 	-44.4//-13.4 & 2.5  & 	30.6	&  	9.1 & 2.9 \\
\hline
 {\cellcolor{green!25}} Theia & 0.6  & 37.9	& 	-68.4//-45.4 		& 1.3  & 	36.6	&  		15.1 & 1.5 \\
\hline
 {\cellcolor{green!25}} SN & 1  & 42.4			& 	-33.4//-8.5 & 0.8  & 	41	&  		19.6 & 0.9 \\
\hline
 {\cellcolor{green!25}} NSS & 0.45  & 35.4	& 	-51.3//-31.3 			& 1.7  & 	34	&  		12.6 & 1.9 \\
\hline
 {\cellcolor{green!25}} Karousel  & 0.6  & 37.9	& 	-23.7//-3.7 		& 1.3  & 	36.6	& 15.1 & 1.5 \\
\hline
\hline
 {\cellcolor{red!25}} SpaceX & 0.45  & 35.4	& -40 & 1.7  & 	34	&  		12.6 & 1.9 \\
\hline
 {\cellcolor{red!25}} OW MEO & 0.45  & 35.4	& 	-31 & 1.7  & 	34	&  		12.6 & 1.9 \\	
\hline
 {\cellcolor{red!25}} OW LEO & 0.45  & 35.4	& 	-40 & 1.7  & 	34	&  		12.6 & 1.9 \\
\hline
 {\cellcolor{red!25}} Kepler & 0.45  & 35.4	& 	-42 & 1.7  & 	34	&  		12.6 & 1.9 \\
\hline
 {\cellcolor{red!25}} Theia & 0.6  & 37.9	& 	-36.9 & 1.3  & 	36.6	&  		15.1 & 1.5 \\
\hline
 {\cellcolor{red!25}} SN & 1  & 42.4	& 	-13.5 & 0.8  & 	41	&  		19.6 & 0.9 \\
\hline
 {\cellcolor{red!25}} NSS & 0.45  & 35.4	& 	-24.3 & 1.7  & 	34	&  		12.6 & 1.9 \\
\hline
 {\cellcolor{red!25}} Karousel  & 0.6  & 37.9	& 	-16.9 & 1.3  & 	36.6	& 15.1 & 1.5 \\
\hline
\hline
 {\cellcolor{yellow!25}} Earth St. & 3.7 & 53.7	&  		-8.5 & 0.2  & 	52.4	&  		30.9 & 0.2\\
\hline
 {\cellcolor{yellow!25}} User Te. & 0.75 & 39.9	&  		-13.5  & 1 &	38.5	&  		17 & 1.2	\\
\hline
\end{tabular}
\label{tab:origngsogs}
\end{center}
\end{table}

\bibliographystyle{IEEEtran}
\bibliography{IEEEabrv, bibliography}

\end{document}